\documentstyle[aps,preprint,pra]{revtex}
\begin{document}
%JM begin preamble
\newcommand{\etal}{{\em et al.}\/}
\newcommand{\IP}{inner polarization}
\newcommand{\IPF}{\IP\ function}
\newcommand{\IPFs}{\IP\ functions}
%\renewcommand{\thetable}{\Roman{table}}
%\newcommand{\auth}[2]{#1 #2, }
%\newcommand{\jcite}[4]{{\it #1} {\bf #2}, #3 (#4)}
%\newcommand{\et}{ and }
%JM if ACS journal, uncomment following lines
%\newcommand{\jcite}[4]{{\it #1} {\bf #4}, {\it #2}, #3}
%\newcommand{\auth}[2]{#2, #1;}
%\newcommand{\edit}[2]{#2, #1, Ed.}
%\newcommand{\twoedit}[4]{#2, #1; #4, #3, Eds.}
%\newcommand{\inpress}[1]{{\it #1}, in press}
%\newcommand{\subm}[1]{{\it #1}, submitted}
%\newcommand{\et}{}
%\newcommand{\twoauth}[4]{#2, #1; #4, #3;}
%\newcommand{\andauth}[2]{#2, #1;}
%\newcommand{\book}[4]{{\it #1}; #2: #3, #4}
%\newcommand{\editedbook}[5]{{\it #1}; #2; #3: #4, #5}
%\newcommand{\inbook}[5]{In {\it #1}; #2; #3: #4, #5}
%\newcommand{\tbp}{to be published}
%JM if AIP journal uncomment following lines instead
\newcommand{\auth}[2]{#1 #2, }
\newcommand{\twoauth}[4]{#1 #2 and #3 #4, }
\newcommand{\andauth}[2]{and #1 #2, }
%\newcommand{\jcite}[4]{#1 {\bf #2}, #3 (#4)}
%JM I really prefer this look
\newcommand{\jcite}[4]{{\it #1} {\bf #2}, #3 (#4)}
\newcommand{\et}{ and }
\newcommand{\book}[4]{{\it #1} (#2, #3, #4)}
\newcommand{\erratum}[3]{\jcite{erratum}{#1}{#2}{#3}}
%JM if Elsevier journal uncomment following lines instead
%\newcommand{\auth}[2]{#1 #2, }
%\newcommand{\twoauth}[4]{#1 #2 and #3 #4, }
%\newcommand{\andauth}[2]{and #1 #2, }
%\newcommand{\jcite}[4]{#1 #2 (#4) #3}
%\newcommand{\et}{ and }
%\newcommand{\erratum}[3]{\jcite{erratum}{#1}{#2}{#3}}
%\newcommand{\jcite}[4]{#1 #2 (#4) #3}
\newcommand{\JCP}[3]{\jcite{J. Chem. Phys.}{#1}{#2}{#3}}
\newcommand{\jms}[3]{\jcite{J. Mol. Spectrosc.}{#1}{#2}{#3}}
\newcommand{\jmsp}[3]{\jcite{J. Mol. Spectrosc.}{#1}{#2}{#3}}
\newcommand{\theochem}[3]{\jcite{J. Mol. Struct. ({\sc theochem})}{#1}{#2}{#3}}
\newcommand{\jmstr}[3]{\jcite{J. Mol. Struct.}{#1}{#2}{#3}}
\newcommand{\cpl}[3]{\jcite{Chem. Phys. Lett.}{#1}{#2}{#3}}
\newcommand{\cp}[3]{\jcite{Chem. Phys.}{#1}{#2}{#3}}
\newcommand{\pr}[3]{\jcite{Phys. Rev.}{#1}{#2}{#3}}
\newcommand{\jpc}[3]{\jcite{J. Phys. Chem.}{#1}{#2}{#3}}
\newcommand{\jpca}[3]{\jcite{J. Phys. Chem. A}{#1}{#2}{#3}}
\newcommand{\jcc}[3]{\jcite{J. Comput. Chem.}{#1}{#2}{#3}}
\newcommand{\molphys}[3]{\jcite{Mol. Phys.}{#1}{#2}{#3}}
\newcommand{\physrev}[3]{\jcite{Phys. Rev.}{#1}{#2}{#3}}
\newcommand{\mph}[3]{\jcite{Mol. Phys.}{#1}{#2}{#3}}
\newcommand{\cpc}[3]{\jcite{Comput. Phys. Commun.}{#1}{#2}{#3}}
\newcommand{\jcsfii}[3]{\jcite{J. Chem. Soc. Faraday Trans. II}{#1}{#2}{#3}}
\newcommand{\jacs}[3]{\jcite{J. Am. Chem. Soc.}{#1}{#2}{#3}}
\newcommand{\ijqcs}[3]{\jcite{Int. J. Quantum Chem. Symp.}{#1}{#2}{#3}}
\newcommand{\ijqc}[3]{\jcite{Int. J. Quantum Chem.}{#1}{#2}{#3}}
\newcommand{\spa}[3]{\jcite{Spectrochim. Acta A}{#1}{#2}{#3}}
\newcommand{\tca}[3]{\jcite{Theor. Chem. Acc.}{#1}{#2}{#3}}
\newcommand{\tcaold}[3]{\jcite{Theor. Chim. Acta}{#1}{#2}{#3}}
\newcommand{\jpcrd}[3]{\jcite{J. Phys. Chem. Ref. Data}{#1}{#2}{#3}}
\newcommand{\APJ}[3]{\jcite{Astrophys. J.}{#1}{#2}{#3}}
\newcommand{\astast}[3]{\jcite{Astron. Astrophys.}{#1}{#2}{#3}}
\newcommand{\arpc}[3]{\jcite{Ann. Rev. Phys. Chem.}{#1}{#2}{#3}}

%JM end preamble

\draft
\title{
Benchmark ab initio thermochemistry of the isomers of diimide, N$_2$H$_2$, 
using accurate computed structures and anharmonic force fields}
\author{Jan M.L. Martin$^*$}
\address{Department of Organic Chemistry,
Kimmelman Building, Room 262,
Weizmann Institute of Science,
76100 Re\d{h}ovot, Israel. {\em Email:} \verb|comartin@wicc.weizmann.ac.il|
}
\author{Peter R. Taylor}
\address{San Diego Supercomputer Center and Department of
Chemistry and Biochemistry, University of California, San Diego, P.O. Box
85608, San Diego, CA 92186-5608, USA. {\em Email:~\verb|taylor@sdsc.edu|}}
\date{MS MP013 for B. Liu memorial issue of {\em Mol. Phys.}; received June 4, 1998;
revised \today}
\maketitle
\begin{abstract}
A benchmark ab initio study on the thermochemistry of the trans-HNNH, cis-HNNH,
and H$_2$NN isomers of diazene has been carried out
using the CCSD(T) coupled cluster method, basis sets as large as
$[7s6p5d4f3g2h/5s4p3d2f1g]$, and extrapolations towards the 1-particle
basis set limit. The effects on inner-shell correlation and of 
anharmonicity in the zero-point energy were taken into account: accurate
geometries and anharmonic force fields were thus obtained as by-products.
Our best computed
$\Delta H^\circ_{f,0}$ for trans-HNNH, 49.2$\pm$0.3 kcal/mol, is in very good
agreement with a recent experimental lower limit of 48.8$\pm$0.5 kcal/mol.
CCSD(T) basis set limit values for the isomerization energies at 0 K are
5.2$\pm$0.2 
kcal/mol (cis-trans) and 24.1$\pm$0.2 kcal/mol (iso-trans). Our best computed
geometry for trans-HNNH, 
$r_e$(NN)=1.2468 \AA,
$r_e$(NH)=1.0283 \AA, and $\theta_e$=106.17$^\circ$, reproduces the 
precisely known ground-state rotational constants of trans-HNNH to within
better than 0.1 \%. 
The rotation-vibration spectra of both cis-HNNH and H$_2$NN are dominated
by very strong Coriolis and Fermi resonances. In addition, the NH stretches
in H$_2$NN are so strongly anharmonic that vibrational perturbation theory
breaks down, and the molecule appears to be an excellent test case for
variational treatments of the vibrational Schr\"odinger equation. 
\end{abstract}

\section{Introduction}

The existence of N$_2$H$_2$ (diazene, diimide) was suggested
as early as 1892 \cite{1892} as an intermediate in the decomposition
of azoformic acid. The trans isomer was finally discovered in 1958
 by mass
spectrometry as a gaseous discharge product of hydrazine\cite{Fon58},
and around the same time in the infrared as a photolysis product of matrix
isolated hydrazoic acid\cite{Pim57,Ros65}. The compound is of importance in
organic chemistry for the stereospecific reduction of olefins\cite{Nob74}
 and of course as the parent of a large number of azo
compounds,
and in inorganic chemistry as a ligand for transition
metal complexes\cite{Vei76}. For comprehensive (if somewhat older)
reviews on the 
preparation, properties, and reactions of diimide, the reader
is referred to Back\cite{Bac84} and to a volume\cite{patai} in the Patai
series on the chemistry of functional groups.

Early work on its rotation-vibration 
spectroscopy was reviewed by Craig and Levin\cite{Cra79} (CL).
Its anharmonic force field was studied ab initio at the MP2/$[4s3p2d1f]$
 and CCSD/$[3s2p1d]$ 
levels by Kobayashi, Bl\'udsky,
Koch, and J{\o}rgenson (KBKJ)\cite{Kob93} --- who also considered the
cis-diazene and isodiazene isomers ---
 and very recently at the CCSD(T)/$[4s3p2d1f]$ 
level by the present authors (MT)
\cite{n2h2}, who also reviewed more recent spectroscopic information. 
After publication of this latter paper,
Demaison, Hegelund, and B\"urger (DHB)\cite{Dem97} published new
experimentally derived equilibrium geometry data, based in part
on a re-analysis of recent high-resolution data
for trans-HNNH \cite{Heg94}, trans-DNND \cite{Heg96}, and 
trans-HNND \cite{Heg97}. 

Experimental information on the cis isomer is quite limited, basically
consisting of tentative assignments of bands at 3074 and 1279 cm$^{-1}$ by 
Rosengren and Pimentel (RP)\cite{Ros65} and of bands at 1034, 1347, 3025, and
3116 cm$^{-1}$ by Wiberg, Fisher, and Bachhuber (WFB)\cite{Wib77}.
CL also derived estimated harmonic frequencies from an empirical force
field based on the trans-HNNH frequencies. In light of the fact that 
some of these frequencies were misassigned\cite{n2h2} and that various
force constants had to be neglected for want of sufficient experimental 
data, these frequencies are of limited reliability.

The isodiazene isomer was predicted about a century ago
to play a key role as an intermediate in the
chemistry of azo compounds\cite{1900}. Following earlier
reporting of the synthesis of the substituted isodiazene
(2,2,6,6-tetramethylpiperidyl)nitrene\cite{Der81},
isolation and characterization by low-temperature matrix
infrared spectroscopy was reported by Sylwester and Derwan (SD)\cite{Syl84}.
%1574 [14N,14N], 1548 [14N,15N] cm-1 in solid argon
Teles et al. (TMHS)\cite{Tel89} obtained isodiazene by 
photolysis of amino-isocyanates in
argon at 12 K, and recorded a complete infrared spectrum which 
they assigned with the help of fairly low-level ab initio 
harmonic frequency calculations. Finally, Goldberg et al.\cite{Gol93}
reported mass
spectrometric detection of isodiazene and its cation
in the gas phase. These authors also carried out G2 level calculations
on the relative energies of various minima and transition states on
the N$_2$H$_2$ potential surface, as did Smith\cite{Smi93}.

Ab initio (e.g.\cite{Jen87,Wal89,Pop91,Gol93,Ang96}) and
density functional (e.g.\cite{And93,Jur96})
calculations on the relative stability of the trans-diazene, cis-diazene, 
and isodiazene isomers have consistently shown a stability ordering
trans $>$ cis $>$ iso, as well as high (in excess of 40 kcal/mol) isomerization
barriers between the isomers. (Very recently, the rigid cis-trans rotation
of the molecule was proposed and studied\cite{Mac98} as a test case for a
new multireference coupled cluster method\cite{Mac98}, the transition state
being a `real-life' alternative for H$_4$ as an essentially perfect 
two-configuration reference problem.)

The heat of formation of trans-HNNH is not very well established. The original paper
by Foner and Hudson reports a mass spectrometric $\Delta H^\circ_{f,0}$=52.4$\pm$2.0 kcal/mol.
By photoionization mass spectrometry, Rusci\'c and Berkowitz\cite{Rus91} established
a lower limit, $\Delta H^\circ_{f,0}\geq 46.6\pm0.8$ kcal/mol, which is consistent
with a calculated value using the G2 model\cite{g2} by Pople and Curtiss\cite{Pop91},
49.6 kcal/mol. The gap between theoretical value and experimental upper limit was 
narrowed by new vacuum UV photolysis experiments by Biehl and Stuhl (BS) \cite{Bie94}
who obtained $\Delta H^\circ_{f,0}\geq 48.8\pm0.5$ kcal/mol. Clearly, the availability
of a benchmark ab initio value would be highly desirable.

We have recently developed basis set extrapolation techniques\cite{l4,c2h4tae} which 
permit the calculation of molecular total atomization energies (TAE$_e$ at the bottom 
of the well, TAE$_0$ at 0 K) with a mean absolute error as low as 0.12 kcal/mol.
Using this technique, we were recently able\cite{bf3} to resolve a long-standing
controversy concerning the heat of formation of B(g).  One objective of the present
work is to obtain a benchmark heat of formation of trans-HNNH, as well as benchmark
values for the isomerization energies in the \{trans,cis,iso\}diazene system.

As by-products of these calculations (which require inclusion of inner-shell
correlation and of anharmonic contributions to the total atomization energy)
we will also present accurate computed geometries and anharmonic force fields,
although the latter are not the primary focus of the present work.

\section{Methods}

All electronic structure calculations were carried out using
the MOLPRO 96.4\cite{molpro} quantum chemistry package
running on DEC Alpha 500/500 and SGI Origin 2000 computers
at the Weizmann Institute, and on the Cray C90 at San Diego
Supercomputer Center.

The CCSD(T)
electron correlation method\cite{Rag89,Wat93}, as implemented
by Hampel \etal\cite{Ham92}, has been used
throughout. (For the atomic calculations involved in the TAE
determinations, the definition of the
restricted open-shell CCSD(T) energy according to
Ref.\cite{Wat93} is used.)
The acronym stands for coupled cluster with all single
and double substitutions\cite{Pur82} augmented by a quasiperturbative
account for triple excitations\cite{Rag89}. From extensive
studies (see \cite{Lee95} for a review) this method is known to yield
correlation energies very close to the exact $n$-particle solution
within the given basis set as long as the Hartree-Fock determinant is
a reasonably good zero-order reference wave function.
The ${\cal T}_1$ diagnostic, proposed\cite{Lee89} as a measure of the
importance of nondynamical correlation, was found to be 0.017 and 0.018,
respectively, for trans-HNNH and cis-HNNH: these values suggest a 
wavefunction dominated by a single reference determinant. The computed
${\cal T}_1$ for isodiazene, 0.030, suggests a mild degree of nondynamical
correlation which experience suggests\cite{Lee95} is still well within
the range of applicability for CCSD(T).

Calculations including only valence correlation were carried out
using the cc-pV$n$Z and aug$'$-cc-pV$n$Z ($n$=D, T, Q, 5) basis sets
of Dunning and coworkers\cite{Dun89,Ken92}. The highest angular momenta
present in these basis sets are \{d,f,g,h\} in the series 
\{D,T,Q,5\}.
The augmented correlation consistent (aug-cc-pV$n$Z) basis sets
differ from the parent cc-pV$n$Z basis set by the addition of
one diffuse (anion) function for each angular momentum. It was
previously established that such basis functions are essential
for accurate computed bond angles\cite{hf} 
and atomization energies\cite{l4,watoc} of strongly polar molecules.
The aug$'$-cc-pV$n$Z notation\cite{Del93} stands for the combination of a regular
cc-pV$n$Z basis set on hydrogen with an aug-cc-pV$n$Z basis set
on first-row atoms. We previously found\cite{l4} that this affects
computed atomization energies by less than 0.1 kcal/mol.
In the interest of brevity, the standard acronyms cc-pV$n$Z
and aug$'$-cc-pV$n$Z will be replaced by V$n$Z and A$'$V$n$Z,
respectively.

The effect of inner-shell correlation was assessed by taking the
difference between CCSD(T) calculations with and without constraining
the inner-shell orbitals to be doubly occupied. Inner-shell correlation
requires sufficient flexibility of the $s$ basis set in the high-exponent
region as well as the presence of high-exponent $p$, $d$, and $f$ functions.
In the present work, we have employed the Martin-Taylor core correlation 
basis set\cite{hf}, which was previously found\cite{cc} to recover essentially
the entire differential effect of inner-shell correlation for first-row
molecules.

Geometry optimizations were carried out by repeated multivariate
parabolic interpolation with a step size of 0.001 $a_0$ or radian, and 
a convergence threshold of about 10$^{-5}$ $a_0$ or radian. Quartic
force fields were set up by finite differentiation
in symmetry-adapted coordinates.
In order to keep higher-order contamination in the quartic portion of
the force field to a minimum, fairly small step
sizes 0.01 \AA\ and radian were used and CCSD(T) energies converged to
essentially machine precision.
Generation of the displaced Cartesian geometries and transformation
of the internal coordinate force field to Cartesian coordinates
were carried out with the help of the INTDER\cite{intder}
program. The anharmonic 
spectroscopic analysis was carried out by standard second-order
rovibrational perturbation theory\cite{Pap81,Wat77} using
a modified version of SPECTRO\cite{spectro,Gaw90}. Cubic and quartic
resonances were accounted for using the method previously described
in MT, as implemented by one of us\cite{polyad}.

\section{Results and discussion}

\subsection{Thermochemistry}

All relevant data can be found in Table \ref{tab:thermo}.  In this
molecule, the SCF and valence correlation contributions to the total
atomization energy (TAE) are clearly equally important.

Using the cc-pV$n$Z basis sets, basis set convergence for the SCF
contribution is atypically slow: enlarging the basis set from cc-pVTZ
to cc-pVQZ affects all three TAE values by around 2 kcal/mol, while
further basis set expansion to cc-pV5Z still contributes another 0.5
kcal/mol, on average. By contrast, convergence in the aug$'$-cc-pV$n$Z
series is considerably faster: improving the basis set from
aug$'$-cc-pVTZ to aug$'$-cc-pVQZ increases TAE by only about 1
kcal/mol, while further enlargement to aug$'$-cc-pV5Z affects all
values by only 0.1--0.2 kcal/mol. It is also worth noting that the
SCF/cc-pVDZ and SCF/aug$'$-cc-pVDZ TAE values differ by 5--5.5
kcal/mol. These observations clearly demonstrates that diffuse
functions on N are essential for a balanced description of these
species.

For the sake of elegance, an extrapolation to the one-particle basis
set limit is required. We have considered two alternatives: an
$A+B/(l+1/2)^{5}$ extrapolation following the suggestion of Petersson
et al.\cite{cbs} and Ref.\cite{c2h4tae}, and the geometric
extrapolation formula originally proposed by Feller\cite{Fel92}.
The extrapolated limits from the V$n$Z and A$'$V$n$Z series differ
appreciably with the former formula: the latter yields essentially
identical results for both molecules. In order to clarify this matter
further, we have considered extrapolated total SCF energies for a
number of systems for which numerical Hartree-Fock limits are
available. These data are given in Table \ref{tab:nhf}. It is clearly
seen that the geometric formula most closely reproduces the numerical
SCF energies, although the value of an extrapolation with basis sets
as large as those considered here is largely cosmetic.

Following the pioneering work by Schwartz\cite{Sch63}, Hill\cite{Hil85}
and Kutzelnigg and Morgan\cite{Kut92} showed that the basis set
convergence of pair correlation energies can be expanded as an
asymptotic series in $1/(l+1/2)$, with $l$ the highest angular
momentum appearing in the basis set. Based hereon, Martin\cite{l4}
proposed the use of a 3-point extrapolation formula of the form
$A+B/(l+1/2)^C$. In combination with appropriate SCF extrapolations and
accounts for inner-shell correlation, the present authors\cite{c2h4tae}
found that
the very precisely known experimental TAEs of 15 small polyatomic molecules
could be reproduced to within 0.12 kcal/mol.

In the present case, the A$'$V$n$Z basis sets systematically recover a
slightly larger percentage of the correlation energy than their V$n$Z
counterparts: however, the differences are much smaller than for the
SCF portion of TAE. As we previously found\cite{l4,watoc,c2h4tae} for
molecules with polar bonds, the extrapolated values based on V$n$Z
calculations are substantially higher (about 0.6 kcal/mol in this case)
than those obtained from A$'$V$n$Z results. Relative to the
CCSD(T)/A$'$V5Z result, the extrapolation accounts for 1.5--1.7
kcal/mol.

It is perhaps worth mentioning that the convergence of the {\em sum}
of SCF and correlation energies for relatively small basis sets would
be dominated to a substantial extent by the SCF convergence behavior,
and leads to the erroneous conclusion that overall convergence
behavior is best described by an exponential series.

Inner-shell correlation contributes 0.91, 0.87, and 1.04 kcal/mol,
respectively, for the trans-, cis-, and isodiazene isomers. It thus
leaves the cis-trans equilibrium fundamentally unchanged but does
favor the isodiazene isomer somewhat compared to the other two isomers.

For the `bottom-of-the-well' situation, our best calculations thus
predict a cis-trans isomerization energy of 5.59 kcal/mol and an
iso-trans difference of 24.95 kcal/mol. Inclusion of CCSD(T)/cc-pVQZ
anharmonic zero-point energies considerably affects these values,
favoring cis over trans by 0.38 kcal/mol and isodiazene over trans-diazene
by 0.83 kcal/mol. Our final best isomerization energies at 0 K are then
5.21 kcal/mol (cis-trans) and 24.12 kcal/mol (iso-trans), which we
estimate to be accurate to about 0.1 kcal/mol. The G2 values of
Goldberg et al., 5.0 and 24.1 kcal/mol, are in excellent agreement with
our values.

Combining our computed TAE$_{0}$ of 278.76 kcal/mol for the trans form
with the JANAF heats of formation of N($g$) and H($g$), we obtain a
computed CCSD(T) basis set limit
$\Delta H^{0}_{f,0}$(trans-HNNH)=49.57 kcal/mol, to which we
assign a conservative error bar of $\pm$0.2 kcal/mol. The G2 value of
49.6 kcal/mol\cite{Pop91} is in perfect agreement with our
calibration calculation. The most recent experimental value is a lower limit
of 48.8 $\pm$ 0.5 kcal/mol by Biehl and Stuhl\cite{Bie94}, which is
quite consistent with our calculations.

A litmus test for basis set convergence of our computed value would be
if a 3-point extrapolation from AVQZ, AV5Z, and AV6Z were to yield the
same result as above (AVTZ,AVQZ,AV5Z). While a CCSD(T)/AV6Z
calculation on trans-HNNH is beyond our computational resources, we
can certainly carry out such calculations for the prototype systems
N$_2$ and NH. For these systems, we find differences of -0.13 and
-0.045 kcal/mol, respectively, adding up for HNNH to an estimated
difference of only -0.04 kcal/mol. We can therefore assume that our
result is converged with respect to the 1-particle basis set.

The remaining discrepancy with experiment for N$_2$, then, appears to
be to some extent due to imperfections in the CCSD(T) treatment.
Because their magnitude in trans-HNNH is somewhat hard to quantify, we
affix an overall error bar of about 0.5 kcal/mol to our computed TAE$_e$.

In Ref.\cite{c2h4tae}, it was proposed to add a correction term of
0.126 kcal/mol per bond order involving at least one N atom, which 
greatly improved agreement with experiment for such compounds as N$_2$,
NH$_3$, NNO, and HNO. If the same were done here, this would lead to
an increase of 0.504 kcal/mol in the computed TAE$_0$ to 279.26 kcal/mol,
and a decrease in the computed $\Delta H^\circ_{f,0}$ to 49.07 kcal/mol,
which is within the error bar of the Biehl and Stuhl lower limit.

Alternatively, we may use the 3-parameter empirical correction due to
Martin\cite{Mar92,watoc}
\begin{equation}
\Delta E_{\hbox{correction}}
=a_\sigma n_\sigma + b_\pi n_\pi + (n_\sigma+n_\pi+n_{\hbox{lone pair}})
c_{\hbox{pair}}\label{x}
\end{equation}
in which $n_\sigma,n_\pi,n_{\hbox{lone pair}}$ represent the numbers of $\sigma$
bonds, $\pi$ bonds, and lone pairs, respectively, and the coefficients
$a_\sigma,b_\pi,c_{\hbox{pair}}$ are specific for the basis set, electron
correlation treatment, and (level of theory used for the) reference geometry.
(They are determined by least-squares fitting to a fairly small sample
of very accurately known TAEs.) Ref.\cite{watoc} lists two sets of parameters,
one which only attempts to correct for basis set incompleteness in the 
valence correlation treatment (the inner-shell correlation contribution
being computed explicitly), and another which attempts to absorb the 
inner-shell correlation contribution in the parametrization.
In the present case, as seen in Table \ref{tab:3par}, 
the `implicit core correlation'
set of parameters yields
essentially identical results to the set of parameters
with explicit inclusion of core correlation. The 3-parameter corrected values
converge surprisingly rapidly as a function of basis set
(Table \ref{tab:3par}): from VTZ to A$'$V5Z,
the variation is no larger than about 0.15 kcal/mol. The highest-level value,
A$'$V5Z with explicit core correlation, is TAE$_e$=296.5 kcal/mol, or
TAE$_0$=279.0 kcal/mol, that is $\Delta H^\circ_{f,0}$=49.37 kcal/mol. 
(The mean absolute error for the ``training set'' in Ref.\cite{watoc} is
0.20 kcal/mol at this level.)
Taking the average between the extrapolated and empirically corrected values
and using twice the difference between the values as an estimated error
bar, we finally suggest $\Delta H^\circ_{f,0}$=49.2$\pm$0.3 kcal/mol as our best 
estimate for the heat of formation of trans-diazene.

As expected, basis set convergence for the cis-trans and iso-trans 
isomerization energies is quite fast, and there is no reason why our
computed basis set limits should not be accurate to as little as 0.1 kcal/mol:
we will conservatively double these error bars, leading to final best
computed isomerization energies at 0 K of 5.2$\pm$0.2 (cis-trans)
and 24.1$\pm$0.2 (iso-trans) kcal/mol.

Finally, in order to assess the performance of the "Complete Basis
Set" (CBS) hybrid extrapolation/empirical correction schemes of
Petersson and coworkers\cite{Pet96} for this problem, we have calculated
the TAE of N$_2$N$_2$ and the isomerization energies using the CBS-Q
and CBS-QCI/APNO models\cite{cbs,Pet96} as implemented in GAUSSIAN
94\cite{g94}. The computed $\Delta H^\circ_{f,0}$ values of 50.9 (CBS-Q) and
%
%TAE= 277.4 and 277.5 kcal/mol, respectively
%
50.8 (CBS-QCI/APNO) kcal/mol agree fairly poorly with the present best
estimate: about one-third of the discrepancy (0.61 kcal/mol) is due to
error in the approximate zero-point energy. The fact that G2 theory
apparently agrees better with the benchmark heat of formation for
N$_2$H$_2$ than the CBS-Q and particularly the CBS-QCI/APNO models
goes against the general trend: e.g. for 14 experimentally very precisely
(0.1 kcal/mol or better) known total atomization energies, one of
us\cite{acs} found mean absolute errors of 1.32 kcal/mol for G2 theory,
0.82 kcal/mol for CBS-Q, and 0.45 kcal/mol for CBS-QCI/APNO. (We recall
for comparison that the corresponding error statistics for the best
extrapolation and 3-parameter correction used in the present work are 0.12
and 0.20 kcal/mol, respectively.)
The cis-trans isomerization energy is computed as 5.2 kcal/mol using both
models, in perfect agreement with our best computed value; like for G2
theory, the iso-trans isomerization energy is underestimated, at 23.6
kcal/mol by CBS-Q and 23.7 kcal/mol by the CBS-QCI/APNO model.

\subsection{Geometries}

All relevant data can be found in Table \ref{tab:geo}.
Only for trans-HNNH are experimental geometric data available.
Until recently, these essentially consisted of the older
$r_s$ (substitution) structure of 
Carlotti et al.\cite{Car74}.
Very recently, Demaison, Hegelund, and B\"urger (DHB)\cite{Dem97} 
published a newly
refined $r_z$ geometry: $r_z$(NH)=1.041(1) \AA, $r_z$(NN)=1.252(1) \AA,
and $\theta$=106.3(1)$^\circ$. (For an overview of the definitions of
various experimentally derived bond distances, see the review by
Kuchitsu\cite{Kuc93}.) Using the average of
$r_z-r_e$ corrections obtained
in three different ways (extrapolation from different isotopic values
of $r_z$, CCSD(T)/cc-pVTZ 
calculated rotation-vibration coupling constants from Ref.\cite{n2h2},
and their own MP2/6-311+G(2d,p) harmonic force field), they obtained
the $r_e$ geometry $r_e$(NH)=1.030(1) \AA, $r_e$(NN)=1.247(1) \AA,
and $\theta_e$=106.3$^\circ$. 

We found that the $r_z-r_e$ difference is only weakly affected by basis
set expansion beyond CCSD(T)/cc-pVTZ: at the CCSD(T)/cc-pVQZ level, 
we obtain the correction $r_z-r_e$(NH)=0.01357 \AA, $r_z-r_e$(NN)=0.00550 \AA,
and $\theta_z-\theta_e$=0.061$^\circ$. Applying these to the DHB $r_z$
values, we obtain an $r_e$ geometry that principally differs by $r$(NH)
being about 0.002--0.003 \AA\ shorter. 

Another approach, such as we have followed in previous studies on e.g.
ethylene\cite{c2h4more} and acetylene\cite{c2h2}, consists of obtaining 
a best calculated geometry and computing ground state rotational constants
from it and the anharmonic force field, then comparing the rotational
constants with experiment. For good agreement with experiment in directly
calculated bond distances, inclusion of inner-shell correlation is 
absolutely essential\cite{cc}. In this work, we find that inner-shell
correlation shortens $r$(NN) by 0.0026 \AA\ and $r$(NH) by 0.0013 \AA,
and opens up the NNH bond angle by 0.12$^\circ$. Adding these contributions
in to the CCSD(T)/cc-pVQZ equilibrium geometry, we obtain $r_e$(NN)=1.2468 \AA,
$r_e$(NH)=1.0283 \AA, and $\theta_e$=106.17$^\circ$. From these parameters and
the CCSD(T)/cc-pVQZ force field, we obtain $A_0$, $B_0$, and $C_0$ values
which deviate from experiment by -0.04\%, -0.06 \%, and -0.07 \%, respectively. 
These small discrepancies suggest that our calculated geometry is considerably
closer to experiment than 0.001 \AA\ and 0.1$^\circ$.

However, it could be argued that since trans-HNNH has quite polar bonds, the
use of diffuse function basis sets is in order. And indeed, as can be seen 
from Table \ref{tab:geo}, adding diffuse functions considerably speeds 
up basis set convergence in the bond angle, with CCSD(T)/A$'$VTZ and
CCSD(T)/A$'$VQZ bond angles now only differing by 0.13$^\circ$ (compared to 
0.32$^\circ$ between CCSD(T)/VTZ and CCSD(T)/VQZ). Applying core-correlation
corrections now to the CCSD(T)/A$'$VQZ geometry, we obtain $r_e$(NN)=1.2464 \AA, 
$r_e$(NH)=1.0288 \AA, and  $\theta_e$=106.37$^\circ$. Together with
the CCSD(T)/cc-pVQZ force field, our deviations for the ground-state rotational
constants are then +0.19\%, -0.11\%, and -0.09\%, respectively. Presumably due to
an error compensation, the VQZ+core geometry appears to be the closer to experiment.

Both extrapolated geometries are in agreement about the fact that the DHB value
for $r_e$(NH) is about 0.002 \AA\ too long. Other discrepancies with DHB fall
within the latter's error bars.

Our predicted geometries and ground-state rotational constants 
(Table \ref{tab:geo}) for the 
cis-diazene and isodiazene isomers may facilitate future experimental work 
on these species.

\subsection{Vibrational frequencies}

Computed harmonic frequencies can be found in Table \ref{tab:harm}.
Computed and observed fundamentals are given in Table \ref{tab:fund},
while computed anharmonicities and rotation-vibration coupling constants
are presented in Tables \ref{tab:anhar} and \ref{tab:alpha}, respectively,
together with the relevant resonance constants.

The CCSD(T)/cc-pVQZ anharmonic force fields obtained in the 
present study were principally calculated in order to obtain
reliable zero-point vibrational energies for the thermochemical
calculations. The vibrational spectroscopy of trans-HNNH was discussed in
detail in MT, while too little experimental information
is available for cis-HNNH to make a meaningful comparison possible.

Based on our anharmonic force field calculations, however, we can
draw some qualitative conclusions on the rotation-vibration spectrum of
cis-HNNH. In particular, both $\nu_1$ and $\nu_5$ are involved in
resonance triads, the former with $2\nu_3$ and $2\nu_6$, the latter
with $\nu_2+\nu_6\approx\nu_5$ and $\nu_2+\nu_6\approx\nu_5$. 
The eigenvectors of the former triad include essentially perfect
50:50 mixtures of $\nu_1$ and $2\nu_6$ states:
\begin{equation}
\left(    \begin{array}{c|ccc}
& \left|100000\right> & \left|002000\right> & \left|000002\right>\\
\hline
\left|100000\right> & 3010.07 & -99.398 & -54.045  \\
\left|002000\right> & -91.580 & 2680.61 & 0.910    \\
\left|000002\right> & -65.556 & 0.910  &  3039.32
    \end{array}
\right)\end{equation}
with the eigensolution
\begin{displaymath}
        \begin{array}{l|ccc}
        &  2652.5    &  2986.5   &3091.1 \\
                 \hline
        \left|100000\right> & -0.273&   0.680& -0.680\\
        \left|002000\right> & -0.961&  -0.219&  0.166\\
        \left|000002\right> & -0.036&   0.700&  0.714
           \end{array}\end{displaymath}
Because of the relative positions of the deperturbed $\nu_1^*$ 
(3010.7 cm$^{-1}$)
and $2\nu_6^*$ (3039.3 cm$^{-1}$) band origins, one could assign the 2986.5
cm$^{-1}$ band to $\nu_1$ and the 3091.1 cm$^{-1}$ band to $2\nu_6$; however,
this labeling is somewhat academic.

The resonance matrix involving $\nu_5$ has the structure:
\begin{equation}
\left(    \begin{array}{c|ccc}
& \left|000010\right> & \left|010001\right> & \left|001001\right>\\
\hline
\left|000010\right> & 2920.83 &  21.149 &-114.943  \\
\left|010001\right> &  21.149 & 3061.78 & 1.963    \\
\left|001001\right> &-114.943 & 1.963  &  2852.64
    \end{array}
\right)\end{equation}
with the eigensolution
\begin{displaymath}
        \begin{array}{l|ccc}
        &  2766.2   &  3002.5  &3066.6\\
                 \hline
        \left|000010\right> &  0.600&   0.766& -0.231\\
        \left|010001\right> & -0.048&  -0.254& -0.966\\
        \left|001001\right> &  0.798&  -0.591&  0.115
           \end{array}\end{displaymath}
where we can again label the 2766.2 and 3002.5 cm$^{-1}$ states as
$\nu_3+\nu_6$ and $\nu_5$, respectively, based on the deperturbed band
origins, but the states mix so strongly that the labeling is again
largely meaningless.

A very strong Coriolis resonance ($Z_{46}^a$=7.79 cm$^{-1}$)
between $\nu_4$ and $\nu_6$ is
predicted around the $a$ axis, as is a strong Coriolis resonance around
the $b$ axis between $\nu_3$ and $\nu_4$ ($Z_{34}^b$=-2.40 cm$^{-1}$)
and a weaker one between $\nu_2$ and $\nu_6$ around the $c$ axis
($Z_{26}^c$=0.84 cm$^{-1}$). The quality of the computed Coriolis interaction
parameters can be gauged by comparing CCSD(T)/cc-pVQZ computed with 
experimentally derived values for trans-HNNH. While agreement between the
computed very large $Z_{46}^a$=8.708 cm$^{-1}$ and the experimental value
of Hegelund et al.\cite{Heg94}, 9.18234(58) cm$^{-1}$, is not as good as
one would hope, the more recent value of DHB\cite{Dem97}, 8.5895 cm$^{-1}$,
is actually midway between our CCSD(T)/cc-pVTZ and CCSD(T)/cc-pVQZ 
values. Agreement between computed and observed $Z_{46}^b$ is likewise
quite satisfactory.

Our calculations are rather difficult to reconcile with the assignments
of WFB, as well as with the 1279 cm$^{-1}$ assignment of RP; their 3074
cm$^{-1}$ assignment could correspond to our computed $\nu_2+\nu_6$ band.
Further experimental work on the cis molecule is clearly
required: we hope that our calculations may assist the latter.

In isodiazene, our calculations likewise find four Fermi resonances:
$\nu_1\approx 2\nu_3$, $\nu_1\approx 2\nu_6$, $\nu_2\approx 2\nu_4$,
and $\nu_5\approx\nu_2+\nu_6$. Aside from these strong resonances,
however, the computed vibrational anharmonicities on the deperturbed
$\nu_1$ and $\nu_5$ bands are exceedingly high: 274.5 and 314.8 cm$^{-1}$
at the CCSD(T)/cc-pVQZ level. 
The only case known to the present authors
of a similarly large stretching anharmonicity in a tightly bound
molecule is for the H--N stretch in HNO, for which Lee et al.\cite{hno}
calculated an anharmonic correction of 287.1 cm$^{-1}$ at the same
level of theory as used here. These authors, in comparisons between
variational calculations and 
vibrational perturbation theory found that the latter
essentially breaks down completely
for HNO. It would then stand to reason that the same would occur
for isodiazene, and preliminary variational calculations with the POLYMODE
program\cite{polymode} do indeed suggest such strong mixing that second-order
perturbation theory is fundamentally inappropriate as a treatment. In the
light thereof, it is perhaps not surprising that the agreement between
the computed fundamentals and the matrix IR data of Teles et al. is
atypically poor for this level of theory.
Moreover, we find that the computed harmonic frequencies and vibrational
anharmonicities are unusually sensitive to the basis set. That is, the
deperturbed anharmonic corrections for $\nu_1$ and $\nu_5$ at the CCSD(T)/VTZ
level are 278.4 and 335.8 cm$^{-1}$ (the latter a difference of no less than
20 cm$^{-1}$ with the CCSD(T)/VQZ value!), while the corresponding harmonic
frequencies change by 16 and 25 cm$^{-1}$, respectively. 

In addition we find a very strong
Coriolis resonance $Z_{46}^a$=-11.96 cm$^{-1}$, aside from a weaker one,
$Z_{63}^c$=1.02 cm$^{-1}$. Under these circumstances, it seems almost
certain that a variational treatment based on an approximate kinetic
energy operator in the Watson
Hamiltonian\cite{Wat68} (such as implemented in the 
POLYMODE\cite{polymode,Rom85} program) will be likewise inadequate.
No tetratomic variational code with an exact kinetic energy operator
(e.g.\cite{Bra93,Sch96}) 
is available to the present authors, and since this study principally
concerns thermochemistry, we will not pursue this point further in
the present paper.

\section{Conclusions}

A benchmark ab initio study on the thermochemistry of the trans-HNNH, cis-HNNH,
and H$_2$NN isomers of diazene has been carried out. Our best computed
$\Delta H^\circ_{f,0}$ for trans-HNNH, 49.2$\pm$0.3 kcal/mol, is in very good
agreement with a recent experimental lower limit of 48.8$\pm$0.5 kcal/mol.
CCSD(T) basis set limit values for the isomerization energies, 
including contributions of inner-shell
correlation and anharmonicity in the zero-point energy, are 5.2$\pm$0.2 
kcal/mol (cis-trans) and 24.1$\pm$0.2 kcal/mol (iso-trans). Performance
of more approximate methods such as G2 theory
 and the CBS-Q and CBS-QCI schemes
was assessed in detail for this system. For
extrapolation of the SCF contribution to atomization energies, the
Feller-type exponential extrapolation $A+B.C^{-l}$, rather than
the two-point $A+B/(l+1/2)^5$ extrapolation,  appears to be
the formula of choice.

Our best computed
geometry for trans-HNNH, 
$r_e$(NN)=1.2468 \AA,
$r_e$(NH)=1.0283 \AA, and $\theta_e$=106.17$^\circ$, reproduces the 
precisely known ground-state rotational constants of trans-HNNH to within
better than 0.1 \%. We conclude that the NH bond distance in the recent
experimental $r_e$ geometry of Demaison et al.\cite{Dem97} is about 0.002 \AA\ 
too long. 

The rotation-vibration spectra of both cis-HNNH and H$_2$NN are predicted to
have very strong Coriolis and Fermi resonances. In addition, the NH stretches
in H$_2$NN are so strongly anharmonic that vibrational perturbation theory
breaks down, and the molecule appears to be an excellent test case for
variational treatments of the vibrational Schr\"odinger equation. 

\acknowledgments

JM is a Yigal Allon Fellow, the incumbent of the Helen and Milton
A. Kimmelman Career Development Chair (Weizmann Institute), and
an Honorary Research Associate (``Onderzoeksleider
in eremandaat'') of the
National Science Foundation of Belgium (NFWO/FNRS).
This research was supported by the National
Science Foundation (USA) through Cooperative Agreement DACI-9619020 and by
Grant No. CHE-9700627~(PRT), and by a grant of computer time from SDSC.

\begin{table}
\caption{Convergence of different components (kcal/mol)
of the total atomization
energy of trans-HNNH and the cis-trans and iso-trans isomerization
energies\label{tab:thermo}}
\squeezetable
\begin{tabular}{lcccccccc}
 & TAE(trans-HNNH) & $\Delta E$(cis-trans) & $\Delta E$(iso-trans)\\
\hline
\multicolumn{3}{c}{SCF contribution}\\
\hline
SCF/VDZ                & 143.18  &  5.78    &  18.07 \\
SCF/VTZ                & 152.61  &  5.88    &  17.83 \\
SCF/VQZ                & 154.63  &  5.98    &  17.96 \\
SCF/V5Z                & 155.15  &  6.07    &  17.96 \\
SCF/V$\infty$Z (a)     & 155.29  &  6.10    &  17.94 \\
SCF/V$\infty$Z (b)     & 155.46  &  6.12    &  17.95 \\
SCF/A$'$VDZ            & 148.63  &  6.18    &  18.20 \\
SCF/A$'$VTZ            & 154.05  &  6.16    &  17.91 \\
SCF/A$'$VQZ            & 155.26  &  6.18    &  18.04 \\
SCF/A$'$V5Z            & 155.30  &  6.09    &  17.99 \\
SCF/A$'$V$\infty$Z (a) & 155.30  &  6.08    &  17.98 \\
SCF/A$'$V$\infty$Z (b) & 155.33  &  6.04    &  17.96 \\
\hline
\multicolumn{3}{c}{Valence correlation contribution}\\
\hline
CCSD(T)-SCF/VDZ        & 115.64  & -0.70    &  9.05 \\
CCSD(T)-SCF/VTZ        & 129.40  & -0.74    &  7.89 \\
CCSD(T)-SCF/VQZ        & 135.57  & -0.77    &  7.15 \\
CCSD(T)-SCF/V5Z        & 137.94  & -0.74    &  6.98 \\
CCSD(T)-SCF/V$\infty$Z (c)&140.56& -0.63    &  6.99 \\
CCSD(T)-SCF/A$'$VDZ    & 116.40  & -0.86    &  7.92 \\
CCSD(T)-SCF/A$'$VTZ    & 130.96  & -0.79    &  7.24 \\
CCSD(T)-SCF/A$'$VQZ    & 136.44  & -0.79    &  6.86 \\
CCSD(T)-SCF/A$'$V5Z    & 138.33  & -0.72    &  6.88 \\
CCSD(T)-SCF/A$'$V$\infty$Z (c)&140.05& -0.53&  7.10 \\
\hline
\multicolumn{3}{c}{Other contributions}\\
\hline
CCSD(T)/MTcore-MTnocore& 0.908   & 0.039    & -0.127 \\
best at bottom of well & 296.26  & 5.59     & 24.95 \\
$\Delta$ZPE            &-17.527  &-0.38     &-0.83 \\
best at 0 K            & 278.73  & 5.21     & 24.12 \\
\end{tabular}

(a) Using geometric extrapolation\cite{Fel92}, $A+B.C^{-l}$, from SCF
components of TAE$_e$ with three largest basis sets in series

(b) Using extrapolation\cite{cbs} $A+B/(l+1/2)^5$ from  SCF
components of TAE$_e$ with two largest basis sets in series

(c) Using variable-exponent $l$-extrapolation\cite{l4}, $A+B/(l+1/2)^C$,
from correlation components of TAE$_e$ 
for three largest basis sets in series

\end{table}

\vspace*{1in}

\begin{table}
\caption{Comparison of performance for SCF basis set extrapolations. All energies in hartree\label{tab:nhf}}
\begin{tabular}{lccc}
   & numerical HF$^a$  &    Feller(Q56)$^b$   &   Schwartz5(56)$^c$ \\
\hline
Ne                   &-128.54709809   & -128.547089   &-128.547284\\
N$_2$($R$=2.068 $a_0$) &-108.9938257    & -108.993818   &-108.993988\\
BH($R$=2.336 $a_0$)    & -25.1315987    &  -25.131601   & -25.131629\\
H$_2$($R$=1.4 $a_0$)   &  -1.13362957   &   -1.133625   &  -1.133634\\
H                    &  -0.5 exactly  &   -0.500000     &  -0.500003\\
BF($R$=2.386 $a_0$)$^d$ & -124.1687792  &   -124.16875956 &   -124.168904 \\
\end{tabular}

(a) D. Moncrieff and S. Wilson, \jcite{Mol. Phys.}{85}{103}{1995};
J. Kobus, D. Moncrieff, and S. Wilson, \jcite{Mol. Phys.}{86}{1315}{1995}.
Bond distances $R$ taken from these references

(b) geometric extrapolation $A+B.C^{-l}$ from SCF/cc-pVQZ, SCF/cc-pV5Z,
and SCF/cc-pV6Z energies 

(c) 2-point extrapolation $A+B/(l+1/2)^5$ from SCF/cc-pV5Z
and SCF/cc-pV6Z energies  

(d) aug-cc-pV$n$Z basis sets used\cite{bf3}
\end{table}

\begin{table}
\caption{Basis set convergence of the 3-parameter corrected
TAE$_e$ of trans-HNNH (kcal/mol)\label{tab:3par}}
\begin{tabular}{lcc}
& implicit core corr.$^a$ & explicit core corr.$^b$\\
\hline
CCSD(T)/VDZ     &   297.96 & 297.97\\
CCSD(T)/VTZ     &   296.53 & 296.55\\
CCSD(T)/VQZ     &   296.44 & 296.45\\
CCSD(T)/V5Z     &   296.50 & 296.51\\
CCSD(T)/A$'$VDZ &   297.87 & 297.88\\
CCSD(T)/A$'$VTZ &   296.59 & 296.60\\
CCSD(T)/A$'$VQZ &   296.50 & 296.51\\
CCSD(T)/A$'$V5Z &   296.49 & 296.50\\
$l$-extrapolated$^c$& 296.26 & 296.26\\
\end{tabular}

(a) core correlation absorbed in the parametrization of the
correction 

(b) core correlation contribution
computed explicitly as difference between
CCSD(T)/MTcore and CCSD(T)/MTnocore

(c) See Table \ref{tab:thermo}

\end{table}

\begin{table}
\caption{Convergence of CCSD(T) computed $r_e$ geometries (\AA, degrees),
best computed $r_z$ and $r_g$ geometries (\AA, degrees)
and best computed and observed ground-state rotational constants (cm$^{-1}$)
\label{tab:geo}}
\begin{tabular}{lccccccccc}
 & trans &  &  & cis &  &  & iso &  & \\
 & $r_e$(NN) & $r_e$(NH) & $\theta_e$(NH) & $r_e$(NN) & $r_e$(NH) &
 $\theta_e$(NH)& $r_e$(NN) & $r_e$(NH) & $\theta_e$(NH)\\
\hline
VDZ & 1.2643 & 1.0447 & 104.97 & 1.2592 & 1.0501 & 111.48 & 1.2280 & 1.0528 &
124.46\\
VTZ & 1.2536 & 1.0310 & 105.73 & 1.2512 & 1.0360 & 111.64 & 1.2214 & 1.0370 &
123.70\\
VQZ & 1.2494 & 1.0294 & 106.05 & 1.2481 & 1.0343 & 111.79 & 1.2194 & 1.0351 &
123.49\\
A'VDZ & 1.2660 & 1.0423 & 105.74 & 1.2652 & 1.0463 & 111.53 & 1.2373 & 1.0443 &
123.50\\
A'VTZ & 1.2529 & 1.0320 & 106.12 & 1.2525 & 1.0367 & 111.77 & 1.2242 & 1.0361 &
123.39\\
A'VQZ & 1.2489 & 1.0300 & 106.25 & 1.2486 & 1.0347 & 111.85 & 1.2206 & 1.0350 &
123.36\\
MTcore & 1.2471 & 1.0285 & 105.97 & 1.2450 & 1.0334 & 111.83 & 1.2153 & 1.0351 &
123.70\\
MTnocore & 1.2497 & 1.0297 & 105.85 & 1.2475 & 1.0346 & 111.75 & 1.2175 & 1.0360
& 123.71\\
best calc.$^a$& 1.2468 & 1.0281 & 106.17 & 1.2456 & 1.0331 & 111.88 & 1.2172 & 1.0342
& 123.49\\
Expt.\cite{Dem97} & 1.247(1) & 1.030(1) & 106.3(1) &---&---&---&---&---&---\\
\hline
 & $A_e$ & $B_e$ & $C_e$ & $A_e$ & $B_e$ & $C_e$ & $A_e$ & $B_e$ & $C_e$ \\
best calc.$^a$ & 10.12689 &1.31151 &1.31151 &9.75433 &1.30580 &1.15163
 &11.24253&1.29790&1.16357 \\
 & $A_0$ & $B_0$ & $C_0$ & $A_0$ & $B_0$ & $C_0$ & $A_0$ & $B_0$ & $C_0$ \\
best calc.$^a$ & 10.00064 &1.30354 &1.14917 &9.65724 &1.29706 &1.13945
&11.06453&1.29487&1.15518\\
expt.\cite{Heg94} & 10.001203(5) & 1.3043373(6) & 1.1499757(6) & --- & --- & ---
& --- & --- & --- \\
\hline
 & $r_z$(NN) & $r_z$(NH) & $\theta_z$(NH) & $r_z$(NN) & $r_z$(NH) &
 $\theta_z$(NH)& $r_z$(NN) & $r_z$(NH) & $\theta_z$(NH)\\
best calc.$^a$ & 1.2523 &1.0418  &106.23  &1.2503  &1.0471  &112.33  &1.2194 &1.0498 &123.59 \\
Expt.\cite{Dem97} & 1.252(1) & 1.041(1) & 106.3(1) &---&---&---&---&---&---\\
 & $r_g$(NN) & $r_g$(NH) & $\theta_g$(NH) & $r_g$(NN) & $r_g$(NH) &
 $\theta_g$(NH)& $r_g$(NN) & $r_g$(NH) & $\theta_g$(NH)\\
 & 1.2524  & 1.0507  & 106.18  & 1.2506  & 1.0561  & 112.18  & 1.2198  & 1.0596 
 & 123.28  \\ 
\end{tabular}

(a) CCSD(T)/cc-pVQZ+CCSD(T)/MTcore-CCSD(T)/MTnocore
\end{table}

\begin{table}
\caption{Basis set convergence of CCSD(T) computed harmonic frequencies
(cm$^{-1}$) for isomers of diazene.\label{tab:harm}}
\begin{tabular}{lcccccccc}
 & VDZ & VTZ & VQZ & A'VDZ & A'VTZ & A'VQZ & MTcore & MTnocore\\
trans-diazene &  &  &  &  &  &  &  & \\
\hline
$\omega_1$ ($a_g$) & 3281.7 & 3269.8 & 3278.3 & 3242.4 & 3264.2 & 3276.9 & 3276.7 &
3268.9\\
$\omega_2$ ($a_g$) & 1614.8 & 1621.8 & 1619.9 & 1608.4 & 1612.3 & 1616.3 & 1625.3 &
1623.7\\
$\omega_3$ ($a_g$) & 1569.4 & 1558.4 & 1567.3 & 1550.0 & 1552.5 & 1564.8 & 1566.7 &
1559.4\\
$\omega_4$ ($a_u$) & 1317.5 & 1328.4 & 1327.7 & 1304.6 & 1319.2 & 1323.2 & 1334.2 &
1331.8\\
$\omega_5$ ($b_u$) & 3248.9 & 3301.6 & 3310.1 & 3271.4 & 3296.8 & 3309.1 & 3309.2 &
3301.7\\
$\omega_6$ ($b_u$) & 1343.0 & 1350.3 & 1349.8 & 1341.8 & 1345.1 & 1348.2 & 1355.5 &
1353.2\\
\hline
cis-diazene &  &  &  &  &  &  &  & \\
\hline
$\omega_1$ ($a_1$) & 3185.2 & 3235.9 & 3245.4 & 3205.9 & 3231.9 & 3243.8 &  & \\
$\omega_2$ ($a_1$) & 1588.3 & 1575.7 & 1582.7 & 1556.9 & 1564.9 & 1578.2 &  & \\
$\omega_3$ ($a_1$) & 1347.1 & 1371.6 & 1370.1 & 1356.7 & 1364.6 & 1367.7 &  & \\
$\omega_4$ ($a_2$) & 1247.8 & 1258.7 & 1261.8 & 1237.4 & 1251.7 & 1258.7 &  & \\
$\omega_5$ ($b_2$) & 3085.1 & 3146.5 & 3160.7 & 3118.6 & 3147.3 & 3160.9 &  & \\
$\omega_6$ ($b_2$) & 1560.7 & 1565.3 & 1562.9 & 1548.1 & 1556.1 & 1559.8 &  & \\
\hline
isodiazene &  &  &  &  &  & \\
\hline
$\omega_1$ ($a_1$) & 3028.1 & 3107.4 & 3123.3 & 3103.6 & 3124.7 & 3129.1\\
$\omega_2$ ($a_1$) & 1726.0 & 1728.4 & 1728.9 & 1721.7 & 1725.2 & 1727.4\\
$\omega_3$ ($a_1$) & 1602.4 & 1584.6 & 1588.9 & 1571.0 & 1574.1 & 1585.0\\
$\omega_4$ ($b_1$) & 972.2 & 995.6 & 999.1 & 969.6 & 991.2 & 997.0\\
$\omega_5$ ($b_2$) & 2990.8 & 3103.5 & 3128.9 & 3120.0 & 3135.8 & 3141.0\\
$\omega_6$ ($b_2$) & 1316.0 & 1323.0 & 1326.9 & 1302.3 & 1317.2 & 1324.6\\
\end{tabular}
\end{table}

\begin{table}
\caption{Computed and observed fundamentals (cm$^{-1}$) for isomers of
HNNH\label{tab:fund}}
\squeezetable
\begin{tabular}{lccc}
& trans-HNNH & cis-HNNH & H$_2$NN\\
\hline
\multicolumn{4}{c}{CCSD(T)/cc-pVQZ (this work)}\\
\hline
$\nu_1$   &    3051.0$^a$  &  2986.5     &    2866.5\\
$\nu_2$   &    1578.5      &  1548.4     &    1665.3\\
$\nu_3$   &    1528.2      &  1334.7     &    1560.4\\
$\nu_4$   &    1294.2      &  1231.9     &     991.0\\
$\nu_5$   &    3133.3      &  3002.5     &    2769.7\\
$\nu_6$   &    1317.4      &  1520.7     &    1292.7\\
\hline
\multicolumn{4}{c}{CCSD/cc-pVDZ (Ref.\cite{Kob93})}\\
\hline
$\nu_1$   &   3005  & 2962  &  2838 \\
$\nu_2$   &   1609  & 1622  &  1679 \\
$\nu_3$   &   1584  & 1340  &  1613 \\
$\nu_4$   &   1329  & 1244  &  2724 \\
$\nu_5$   &   3096  & 2947  &  1305 \\
$\nu_6$   &   1310  & 1536  &   977 \\
\hline
\multicolumn{4}{c}{Experiment}\\
\hline
$\nu_1$ &3128$^{b,c}$               &[2966]$^e$ &  2862.0$^g$   \\
$\nu_2$ &1582$^c$                   &[1558]$^e$ &   1644.7$^g$ \\
$\nu_3$ &1529$^c$                   &[1390]$^e$ &  1574$^f$,1574.2$^g$ \\
$\nu_4$ &1286$^c$,1288.64786(4)$^d$ &[1259]$^e$ &  1002.7$^g$\\
$\nu_5$ &3120$^c$,3120.28676(6)$^d$ &[2984]$^e$ &  2804.6$^g$\\
$\nu_6$ &1322$^c$,1316.41214(4)$^d$ &[1439]$^e$ &  1287.5$^g$\\
\end{tabular}

(a) $\nu_2+\nu_3$=3127.9 cm$^{-1}$

(b) clearly $\nu_2+\nu_3$ misassigned to fundamental

(c) Craig and Levin\cite{Cra79} 

(d) Ref.\cite{Heg94}

(e) Estimates of Craig and Levin\cite{Cra79}, based on approximate
force field derived from trans-HNNH
 
(f) A. P. Sylwester and P. B. Dervan, {\it J. Am. Chem. Soc.}
{\bf 106}, 4648 (1984)

\end{table}

\begin{table}
\caption{CCSD(T)/cc-pVTZ and CCSD(T)/cc-pVQZ zero-point energies, 
anharmonic corrections, and  
anharmonicity constants for HNNH isomers. Quantities deperturbed for
resonances are marked by an asterisk. All units
are cm$^{-1}$ unless indicated otherwise.\label{tab:anhar}}
\squeezetable
\begin{tabular}{ldddddd}
& \multicolumn{2}{c}{trans-HNNH} & \multicolumn{2}{c}{cis-HNNH} & \multicolumn{2}{c}{H$_2$NN} \\
& cc-pVTZ & cc-pVQZ & cc-pVTZ & cc-pVQZ & cc-pVTZ & cc-pVQZ \\
\hline
ZPE(kcal/mol)&   17.492 &17.527&   17.101   &17.148&  16.607 &16.699\\
$E_o$        &   20.359 &20.594&   34.341   &32.295&  35.785 &35.263\\
$\omega_1-\nu_1^*$ &   210.6 & 209.7 & 236.5 & 235.3 & 278.4 & 274.5 \\
$\omega_2-\nu_2$   &    42.4 &  41.4 &  35.0 &  34.2 &  56.9 &  55.7 \\
$\omega_3-\nu_3$   &    39.1 &  39.1 &  38.4 &  35.5 &  28.8 &  28.5 \\
$\omega_4-\nu_4$   &    34.2 &  33.5 &  31.1 &  30.0 &  11.2 &   8.1 \\
$\omega_5-\nu_5^*$ &   227.4 & 226.4 & 240.4 & 239.9 & 335.8 & 314.8 \\
$\omega_6-\nu_6$   &    32.8 &  32.5 &  42.7 &  42.2 &  31.2 &  34.3\\
$X_{11}$    &  -47.016  &-46.808& -45.549 & -45.693&  -57.506&-56.621\\
$X_{22}$    &   -1.865  & -2.133& -12.350 & -12.138&   -3.209& -3.279\\
$X_{33}$    &  -11.126  &-10.019&   4.992 &   5.638&  -11.290&-11.061\\
$X_{44}$    &   -6.347  & -6.174&  -3.769 &  -3.899&    2.158&  1.908\\
$X_{55}$    &  -45.556  &-45.332& -46.301 & -46.397&  -79.637&-76.380\\
$X_{66}$    &   -4.813  & -4.704&  -1.047 &  -1.008&    1.437& -0.233\\
$X_{21}$    &  -27.577  &-25.420&   2.181 &   2.222&  -32.983&-33.673\\
$X_{31}$    &   -3.877  & -5.914& -40.740 & -40.264&    8.589&  7.786\\
$X_{32}$    &   -8.636  & -9.773&  -7.165 &  -6.854&   -6.152& -6.059\\
$X_{41}$    &  -13.653  &-13.854& -16.641 & -15.054&   -5.591& -5.666\\
$X_{42}$    &   -4.380  & -4.565& -10.042 &  -9.841&   -7.964& -7.089\\
$X_{43}$    &   -9.650  & -9.169&  -4.170 &  -3.713&  -13.963&-13.415\\
$X_{51}$    & -180.771  &-179.861&-201.169&-200.502& -267.335&-260.605\\
$X_{52}$    &  -32.750  &-29.909&   1.948 &   1.854&  -52.212&-49.283\\
$X_{53}$    &   -6.506  & -9.206& -41.131 & -39.929&   12.772& 12.432\\
$X_{54}$    &  -19.012  &-18.833& -17.851 & -17.555&  -13.165& -1.266\\
$X_{61}$    &   -7.235  & -7.347& -34.364 & -34.197&  -29.518&-30.291\\
$X_{62}$    &   -4.048  & -4.644&  -7.606 &  -7.322&   -1.706& -2.095\\
$X_{63}$    &   -5.045  & -4.127&  -3.501 &  -2.700&   -13.692&-13.412\\
$X_{64}$    &    3.602  &  3.786&   1.638 &   1.826&    9.735&  3.549\\
$X_{65}$    &  -33.615  &-33.765& -37.468 & -38.006&  -33.028&-25.402\\
\end{tabular}

For trans-HNNH, the following resonance constants appear in resonance
polyads involving fundamentals (CCSD(T)/cc-pVQZ) :
%CCSD(T)/cc-pVTZ
%$k_{122}$=58.904, $k_{123}$=124.707, $k_{256}$=-259.029,
%$k_{133}$=35.044, $k_{356}$=-185.420 cm$^{-1}$; 
%$K_{22;23}$=5.066, $K_{23;33}$=18.402,
%$K_{21;31}$=-20.248, $K_{25;35}$=-27.723,
%$K_{24;34}$=1.940, $K_{26;36}$=8.519, $K_{22;33}$=-4.316 cm$^{-1}$.
%CCSD(T)/cc-pVQZ
$k_{122}$=48.367, $k_{123}$=127.771, $k_{256}$=-243.953,
$k_{133}$=47.116, $k_{356}$=-210.321 cm$^{-1}$; 
$K_{22;23}$=7.451, $K_{23;33}$=21.293,
$K_{21;31}$=-24.145, $K_{25;35}$=-31.927,
$K_{24;34}$=2.687, $K_{26;36}$=8.428, $K_{22;33}$=-5.680 cm$^{-1}$.

For cis-HNNH, 
%CCSD(T)/cc-pVTZ
%$k_{133}$=-196.183, $k_{166}$=-106.334, $K_{33;66}$=0.969;
%$k_{256}$=-73.121 cm$^{-1}$; 
%%$K_{55;66}$=-9.849 cm$^{-1}$.?????
%$K_{22;23}$=-13.842; $K_{33;32}$= -8.829; $K_{21;31}$=  5.562;
%$K_{24;34}$= -1.275; $K_{25;35}$=  3.276; $K_{26;36}$=  2.512 
%$k_{256}$=57.190; $k_{356}$=322.072 cm$^{-1}$.
%CCSD(T)/cc-pVQZ
$k_{133}$=-198.795, $k_{166}$=-108.090, $K_{33;66}$=1.820;
$k_{256}$=-59.817 cm$^{-1}$;
%$K_{55;66}$=-9.849 cm$^{-1}$.?????
$K_{22;23}$=-12.588; $K_{33;32}$= -9.784; $K_{21;31}$=  5.516;
$K_{24;34}$= -0.645; $K_{25;35}$=  2.384; $K_{26;36}$=  2.422 
$k_{256}$=59.817; $k_{356}$=325.107 cm$^{-1}$.

For H$_2$NN,  
%CCSD(T)/cc-pVTZ
%$k_{133}$=39.620, $k_{166}$=-146.22, $K_{33;66}$=-5.029;
%$k_{244}$=-101.731, $k_{256}$=-257.815 cm$^{-1}$.
CCSD(T)/cc-pVQZ
$k_{133}$=37.952, $k_{166}$=-147.470, $K_{33;66}$=-5.086;
$k_{244}$=-100.842, $k_{256}$=-262.610 cm$^{-1}$.

\end{table}

\begin{table}
\caption{CCSD(T) computed rotational, rotation-vibration
coupling, and Coriolis coupling constants (cm$^{-1}$)\label{tab:alpha}}
\squeezetable
\begin{tabular}{ldddddd}
            &    \multicolumn{2}{c}{trans-HNNH}      &        \multicolumn{2}{c}{cis-HNNH}    &    H$_2$NN\\
\hline
&cc-pVTZ & cc-pVQZ & cc-pVTZ & cc-pVQZ & cc-pVTZ & cc-pVQZ \\
\hline
$A_e$        & 10.00200  & 10.08275  &    9.66754  & 9.71969  &  11.23667  &  11.22354  \\
$B_e$        &  1.30099  & 1.30712  &    1.29594   & 1.30132  &   1.28780  &  1.29338  \\
$C_e$        &  1.15124  & 1.15712  &    1.14275   & 1.14766  &   1.15538  &  1.15973  \\
$\alpha_{1a}$&  0.21332  & 0.21358  &     0.17466  &  0.17547  &  0.21871  &  0.21673   \\
$\alpha_{2a}$& -0.15339  & -0.14091  &     0.02869  &  0.02918  & -0.05290  &  -0.05192   \\
$\alpha_{3a}$&  0.00250  & -0.01664  &    -0.19817  &  -0.20356  &  0.01040  &  0.00947   \\
$\alpha_{4a}$&  0.14715* & 0.14717*  &     0.14862* &  0.14814* &  0.19226*  & 0.18978* \\
$\alpha_{5a}$&  0.16364  & 0.16447  &     0.16068  &  0.16110  &  0.14159  &  0.14166   \\
$\alpha_{6a}$& -0.11293* & -0.11623*  &    -0.11352* &  -0.11683* & -0.14907*  & -0.15040* \\
$\alpha_{1b}$& -0.00111  & -0.00105  &     0.00010  &  0.00009  & -0.00326  &  -0.00318   \\
$\alpha_{2b}$& -0.00003  & +0.00068  &     0.01187  &  0.01171  & -0.00636  &  -0.00645   \\
$\alpha_{3b}$&  0.01133  & 0.01044  &     0.00557* &  0.00550* &  0.01183  &  0.01173   \\
$\alpha_{4b}$&  0.00291* & 0.00287* &     0.00275* &  0.00282* &  0.00931  &  0.00938   \\
$\alpha_{5b}$& -0.00088  & -0.00082  &    -0.00147  &  -0.00142  & -0.00438  &  -0.00415   \\
$\alpha_{6b}$&  0.00375* & 0.00373* &    -0.00131  &  -0.00133  & -0.00123  &  -0.00128   \\
$\alpha_{1c}$&  0.00106  & 0.00111  &     0.00167  &  0.00168  & -0.00094  &  -0.00075   \\
$\alpha_{2c}$&  0.00437* & 0.00456* &     0.01065* &  0.01060* &  0.00554 &  0.00550  \\
$\alpha_{3c}$&  0.01136* & 0.01105* &     0.00607  &  0.00599  &  0.01080* &  0.01068* \\
$\alpha_{4c}$&  0.00069  & 0.00063  &     0.00023  &  0.00020  & -0.00048  &  -0.00050   \\
$\alpha_{5c}$&  0.00145  & 0.00150  &     0.00092  &  0.00097  & -0.00319  &  -0.00308   \\
$\alpha_{6c}$&  0.00498  & 0.00496  &     0.00480* &  0.00479* &  0.00486* &  0.00483* \\
$Z_{23}^c$   &-0.86135$^a$   & -0.86586$^a$  &    ---       &  ---  &  ---      &  ---  \\
$Z_{46}^a$   & 8.50287$^a$   & 8.70815$^a$  &    7.70561   &  7.79458  &-12.00788  &  -11.96119  \\
$Z_{46}^b$   & 2.35531$^a$   & 2.35803$^a$  &    ---       &  ---  &  ---      &  ---  \\
$Z_{62}^c$   & ---       & ---  &    0.84218   &  0.84676  & ---  &  --- \\
$Z_{43}^b$   & ---       & ---  &   -2.40043   &  -2.40189  &  ---      &  ---  \\
$Z_{63}^c$   & ---       & ---  &    ---       &  ---  &  1.01777  &  1.01576  \\
\end{tabular}

Rotation-vibration coupling constants marked with an asterisk have
had terms deleted that were near-singular due to Coriolis resonance.

(a) Experimental values\cite{Heg94}: $Z_{46}^a$=9.18234(58),
$Z_{46}^b$=-2.3663(34) cm$^{-1}$.  Re-analysis by DHB\cite{Dem97}:
$Z_{46}^a$=8.5895, $Z_{46}^b$=-2.41605(7) cm$^{-1}$. 

\end{table}

\end{document}